**Article title: A Review on Data Fusion in Multimodal Learning Analytics and Educational Data mining**

| Wilson Chango; Pontifical Catholic University of Ecuador; Ecuador; |
|---|
| Juan A. Lara; Madrid Open University; Spain |
| Rebeca Cerezo; University of Ovideo; Spain |
| Cristóbal Romero; DaSCI Institute, University of Cordoba; Spain; <u>cromero@uco.es</u> |

**Abstract**

The new educational models such as Smart Learning environments use of digital and context-aware devices to facilitate the learning process. In this new educational scenario, a huge quantity of multimodal students' data from a variety of different sources can be captured, fused and analyze. It offers to researchers and educators a unique opportunity of being able to discover new knowledge to better understand the learning process and to intervene if necessary. However, it is necessary to apply correctly data fusion approaches and techniques in order to combine various sources of Multimodal Learning Data (MLA). These sources or modalities in MLA include audio, video, electrodermal activity data, eye-tracking, user logs and click-stream data, but also learning artifacts and more natural human signals such as gestures, gaze, speech or writing. This survey introduces data fusion in Learning Analytics (LA) and Educational Data Mining (EDM) and how these data fusion techniques have been applied in Smart Learning. It shows the current state of the art by reviewing the main publications, the main type of fused educational data, and the data fusion approaches and techniques used in EDM/LA, as well as the main open problems, trends and challenges in this specific research area.

## 1. INTRODUCTION

The current, and more than likely post-pandemic, scenario seems to point towards new hybrid, more flexible and technological learning environments that can respond to changing circumstances. In this regard, Blended Learning (BL), Hybrid learning (HL) and Smart Learning (SL) are options that comes up repeatedly:

- **Hybrid Learning (HL)** is an educational approach where some individuals participate in person, and some participate online. Instructors and facilitators teach remote and in-person learners at the same time using technology like video conferencing (Raes, 2022).

- **Blended Learning (BL)** is a style of education in which instructors and facilitators combine in-person instruction with online learning activities. Learners complete some components online and do others in person (Sánchez Ruiz et al., 2021).

- **Smart Learning Environments (SLEs)** are physical environments enriched with digital, context-aware, adaptive devices which aim to achieve more effective, better-quality learning (Chen et al., 2021; Tabuenca et al., 2021). SLEs contain multiple sources of data which, combined together, can offer a better understanding of the educational process.

All these new type of learning environments produce a huge amount of student's data interaction. In the last decade, there were an increasing interest in the analysis and exploitation of large amounts of data produced during the learning process in these new educational environments, which are difficult to analyze manually. In fact, there are two related communities about the same Educational Data Science (EDS) research area (Romero & Ventura, 2020):

- **Educational Data Mining (EDM)** can be defined as the application of Data Mining (DM) techniques to this specific type of dataset that come from educational environments to address important educational questions.

- **Learning Analytics (LA)** can be defined as the measurement, collection, analysis and reporting of data about learners and their contexts, for purposes of understanding and optimising learning and the environments in which it occurs.

Both communities share a common interest in data-intensive approaches to educational research and share the goal of enhancing educational practice. However, LA is more focused on the educational challenge and EDM is more focused on the technological challenge. On the one hand, LA is focused on data-driven decision making and integrating the technical and the social/pedagogical dimensions of learning by applying known predictive models. On the other hand, EDM is generally looking for new patterns in data and developing new algorithms and/or models. Regardless of the differences between the LA and EDM communities, the two areas have significant overlap both in the objectives of investigators as well as in the methods and techniques that are used in the investigation (Romero & Ventura, 2020).

Normally, most of the EDM/LA approaches to analyzing educational data are based on using only one specific data source. However, this means being limited by the data source used, which reflects only part of the reality of the educational process. This is a problem in SLEs which produces a fast quantity of data from different sources that make appropriate the use of data fusion techniques for merging all information to correctly understand the peculiarities of the teaching-learning process occurring in these environments. This idea of combined use of several educational data sources has given rise to Multimodal Learning Analytics (MLA). This approach is based on capturing, integrating, and analyzing different sources of educational data which together provide a holistic understanding of the learning process (Sharma & Giannakos, 2020). During multimodal interaction in education environments, new data collection and sensing technologies are making it possible to capture massive amounts of data about students' activity. These technologies include the logging of computer activities, wearable cameras, wearable sensors, biosensors (e.g., that permit measurements of skin conductivity, heartbeat, and electroencephalography), gesture sensing, infrared imaging, and eye tracking. Such techniques enable researchers to have unprecedented insight into the minute-by-minute students' activities, especially those involving multiple dimensions of activity and social interaction (Blikstein & Worsley, 2016). The combination of multimodal data treatment techniques and the intersection with EDM (Educational Data Mining) and LA (Learning Analytics) has been shown to be a productive line of study in recent years (Budaher et al., 2020) (Kaur et al., 2019; Lahat et al., 2015; Poria et al., 2017; Wang et al., 2018). For example, Mitri et al. (2019) proposed a mechanism allowing annotation of multimodal data for subsequent analysis, and Järvelä et al. (2021) gave many examples of the advantages offered by multimodal data with regard to self-regulated learning. Despite numerous advantages, combined use of educational data is not easy, and there are notable challenges, such as differing granularity or the need to align the different timescales for the data collected from different sources.

Data Fusion can be defined as the process of effectively combining data from different sources so that using that data in combination produces more information than each of the sources would separately. In SLEs, this idea has been used to try and exploit multimodal data and better understand the educational process. The general approach of Multimodal Learning Data Fusion and Mining in Smart Classroom is shown in Figure 1. Multimodal data come from different educational environment such as traditional classroom, e-learning or blended and hybrid learning, and different sources or data type. The fusion point and the used fusion technique depend on the educational problem to solve and the DM/LA objective. Finally, new knowledge can be discovered after applying this process for improving our smart classroom.

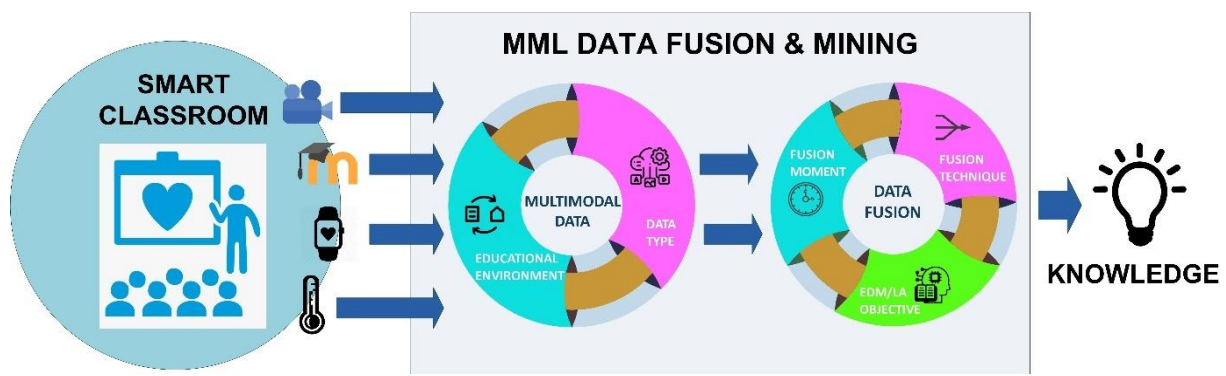

**FIGURE 1. General multimodal data fusion approach for EMD/LA.**

In recent years, there have been an increasing number of survey papers about multimodal educational data (Blikstein & Worsley, 2016; Ochoa, 2017; Shankar et al., 2018). These works examined the application of EDM/LA in multimodal educational data, but which barely touched on data fusion, focusing instead on complex learning tasks (Blikstein & Worsley, 2016), the study of learning analytics architectures (Shankar et al., 2018), and the study of learning environments where multimodal LA is usually applied (Ochoa, 2017). There are also a few review papers more focused in the specific application of data fusion in EDM/LA (Dewan et al., 2019; Han et al., 2020; Nandi et al., 2020). However, they only focused on some specific aspects, including emotion recognition (Nandi et al., 2020), engagement detection (Dewan et al., 2019), or sentiment analysis (Han et al., 2020). Finally, the survey that is most closely related with our current review is from Mu et al. (2020), which focused only on LA, without examining EDM bibliography. That survey only presented the papers analyzed quantitatively, without deeper analysis of the different studies and without establishing the challenges and lines of future study for researchers in this area. So, it is clear that these previous existed and related reviews give an incomplete picture, meaning that there is a need for an up-to-date, comprehensive review of the literature on studies about the specific use of data fusion techniques in SLEs for the application both in LA and EDM. Our objective in this review is to provide a in depth analyses of all the multimodal data used (types, capture methods, etc.), a description of all the data fusion methods and techniques used, the LA and EDM objectives and successful applications, and to identify a set of open challenges and problems. In this way, we will provide the scientific community with a thorough, up-to-date understanding of the current state of this discipline.

We followed the systematic literature review procedure proposed by Tranfield et al. (2003). We used Google Scholar, Web of Science, and Scopus search engines to search for academic papers up to December 2021. In our search we used the following search terms: "Data Fusion" AND ("Multimodal Learning" OR " E-learning" OR "Online learning" OR "Web-based learning" OR "Blended Learning" OR "Hybrid learning" OR Smart Learning" OR "Education"). This preliminary search identified 56 papers whose titles or abstracts included the defined keywords. Then, the papers were selected by reading both the full content of the papers initially downloaded from the search and applying the following inclusion and exclusion rule. We only considered studies in which there was a real educational data fusion process with the aim of applying LA or EDM techniques. It did not include studies which merely used multimodal data from different sources separately, such as Järvelä et al., (2021), nor studies that did use fusion of educational data but without the aim of applying LA/EDM techniques. In this way, we finally selected only 31 papers (20 journal papers and 11 conference papers) published between 2015 and 2021, which confirms the relatively novel nature of this topic.

The remainder of this article is organized as follows: Section 2 provides an analysis of the selected studies according to the type of fused multimodal data; section 3 analyzes those studies from the perspective of the fusion approach or technique used; and finally, section 4 presents our conclusions and outlines the identified challenges and open problems in this area.

## 2. MULTIMODAL DATA

In this section, we analyze what are the most common data used when fusing MLA data. We have differentiated two different viewpoints or fundamental aspects: the educational environment used (in-person, online, and hybrid/blended) and the type of data fused.

In the next two subsections, we have analyzed the previously selected papers. For each source of fused data identified, we show its name, type (audio, video, numerical, etc.), method of capture (camera, microphone, log, etc.) and category. The category will be analyzed using the classification from Mu et al. (2020), summarized in Figure 2, which establishes five different categories of data: digital, physical, physiological, psychometric, and environmental. Digital space referred to various digital traces generated on the system platform during the learning process, such as an online learning platform, virtual experiment platform, or STEAM educational software. Physical space was about the data obtained by various sensors, such as gesture, posture, and body movement. Physiological space referred to the data related to human internal physiological reflection, including EEG and ECG, which objectively reflected students' learning status. In contrast, psychometric space, a relatively common source of learning data, referred to various self-report questionnaires that subjectively reflected the learner's mental state. Environmental space referred to the data about a learning environment where a learner was physically located, such as temperature and weather.

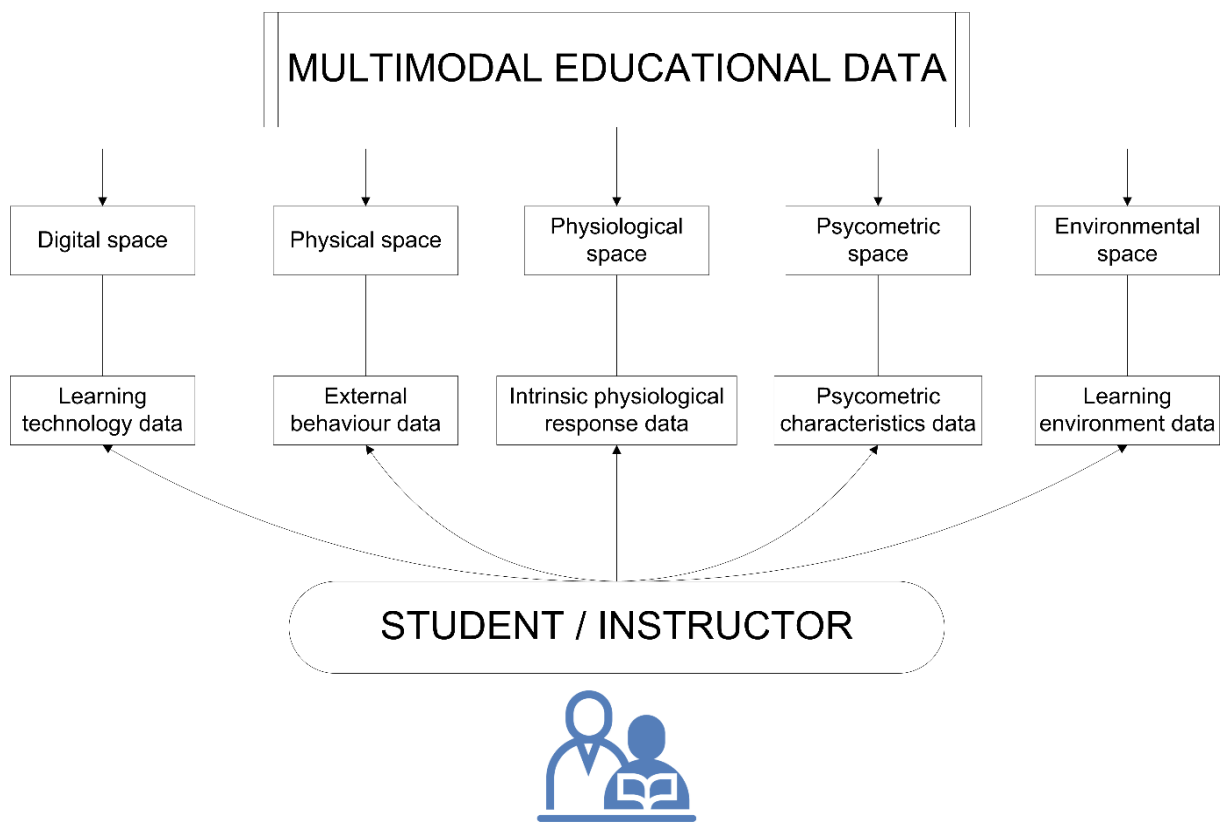

**FIGURE 2. Categories of multimodal educational data, based on Mu et al., 2020.**

## 2.1 Traditional In-person Classroom data

This section presents the different data source from face-to-face traditional teaching. With classroom-based learning, students go to a physical classroom where the teaching and much of the learning takes place. Table 1, show the papers that used these types of data for fusing them. It presents the reference in the first column, the different sources of fused data in the second, and for each source, the type and category (according to the taxonomy in Figure 2), and finally the capture method (Webcam, SMI Eye Tracking Glasses, Electrode, Different Sensors, CSV files and Platform).

**TABLE 1. Sources of multimodal data from in-person classroom.**

| Paper | Source | Type | Category | Capture method |
|---|---|---|---|---|
| Giannakos et al., 2019. | (Student) heart rate | Time Series | Physical | Sensor |
| | Electrodermal activity | Time Series | Physiological | Sensor |
| | Body temperature | Time Series | Physiological | Sensor |
| | Blood volume | Time Series | Physical | Sensor |
| | Electroencephalogram | Numerical | Physiological | Electrode |
| | Eye tracking | Video | Physiological | Webcam |
| Daoudi et al., 2021. | Real-time video recordings | Video | Physical | Webcam |
| | Exchanged messages during playing | Text | Digital | CSV |
| Gadaley et al., 2020. | Student attention to a video source | Video | Physical | Webcam |
| | Student head position to determine attention | Time Series | Physical | Webcam |
| Olsen et al., 2020. | Student audio | Audio | Physical | Webcam |
| | Eye tracking | Video | Physical | Webcam |
| | Questionnaire type test | Numerical | Digital | Platform |
| | Cognitive load by gaze analysis | Time Series | Physical | Sensor |
| | Dialog between students | Text | Digital | Platform Log |
| Mao et al., 2019. | Student images | Photographs | Physical | Webcam |
| Prieto et al., 2018. | Teacher eye-tracking | Video | Physical | SMI Glasses |
| | Teacher movement in the classroom | Time Series | Physical | Sensor |
| | Teacher's presentation | Audio | Physical | SMI Glasses |
| | Teacher's video lessons | Video | Physical | SMI Glasses |
| Worsley, 2014 | Oral test for student | Audio | Physical | Webcam |
| | Student behavior | Video | Physical | Webcam |
| | Hand movements | Time Series | Physical | Sensor |
| Henderson et al., 2020. | Student posture (Kinect) | Time Series | Physical | Sensor |
| | Following movement | Time Series | Physical | Sensor |
| | Interaction between students | Video | Physical | Webcam |
| | Student actions | Text | Digital | Platform |
| Henderson et al., 2019a. | Student posture (Kinect) | Time Series | Physical | Sensor |
| | Following movement | Time Series | Physical | Sensor |
| | Interaction between students | Video | Physical | Webcam |
| | Recordings to identify student frustration | Video | Physical | Webcam |
| | Students' hands (temperature, electrodermal activity and 3D coordinates) | Time Series | Physical/ Physiological | Sensor |
| Ma, et al., 2015. | Recording of the class | Video | Physical | Webcam |
| | Interaction between students and teacher | Video | Physical | Webcam |
| | User and course information | Text | Digital | Plataform |
| | Session start time | Numerical | Digital | Plataform |
| | Browser history | Text | Digital | Plataform |
| | Interviews with students | Text | Digital | Logs |

| Andrade et al., 2016. | Gaze tracking | Time Series | Physical | Sensor |
| | Student gestures | Video | Physical | Webcam |
| Monkaresi et al. 2017. | Writing activity | Text | Digital | Platform |
| | Facial expressions | Video | Physical | Webcam |
| | Hear rate | Time series | Physiological | Sensor |

As the table shows, there is a wide variety of data sources to fuse in face-to-face teaching. In Giannakos et al., 2019, different physical and physiological student data was fused from multiple sensors, including heart rate, body temperature, and blood volume. Some studies used fusion of video and text data (Daoudi et al., 2021), while others fused data from recordings made using 180-degree video cameras (Gadaley et al., 2020). Some of the studies used many different types of sources (Olsen et al., 2020), whereas others focused on specific types of data, such as images in Mao et al. (2019). Most of the studies focused on data collected from students, but some, such as Prieto et al. (2018), used data gathered from wearables worn by the teacher. It is also interesting to note the fusion of multimedia data (audio and video) together with data from students' hands while they performed certain tasks (Worsley, 2014). The only researchers who contributed with several papers to this survey were Henderson et al., whose studies focused on student posture and movement along with other data that varied from one study to the next (Henderson et al., 2020; Henderson et al., 2019a). It was also interesting to see the fusion between physical and digital data gathered via webcam and the learning platform in Ma et al. (2015). The study by Andrade et al. (2016) stood out as it included the analysis of children's gazes and gestures during certain learning activities. Finally, automated detection of student's engagement has done by fusing information from writing activity, videos of their faces after the activity and heart rate (Monkaresi et al. 2017).

## 2.2 Online Classroom data

This section presents the different data sources from online learning. Online education uses the Internet and information and communications technology (ITC) to provide students with tools like chats, blogs, video conferences and shared documents. Table 2 shows the papers that used these types of data for fusing them, using the same column structure as in the previous section.

**TABLE 2. Sources of multimodal data in online settings.**

| Study | Source | Type | Category | Capture method |
| --- | --- | --- | --- | --- |
| Wu et al., 2020. | Teacher gestures (indicative, descriptive, or rhythmical) | Video | Physical | Webcam |
| | Teacher behavior (writing on the board, asking questions, demonstrating, instructing, describing, and non-gesture behavior) | Video | Physical | Webcam |
| | Teacher body movement | Time Series | Physical | Sensor |
| Brodny, 2017. | Facial expressions | Video | Physical | Webcam |
| | Self-report (key-presses and mouse movement patterns) | Video | Physical | Webcam |
| | Physiological signals | Video | Physiological | Webcam |

| | Moodle course (activities, questions, and forum) | Numerical | Digital | Platform |
|---|---|---|---|---|
| Peng & Nagao, 2021. | Student heart rate | Time Series | Physical | Sensor |
| | Conversations with the teacher | Text | Digital | Microphone |
| | Students' mental states | Video | Physical | Webcam |
| Luo et al., 2020. | Head position to measure cognitive attention | Video | Physical | Webcam |
| | Facial expressions (smiles) | Video | Physical | Webcam |
| | Student thoughts | Text | Digital | Platform |
| Henderson et al., 2019b. | Student posture (Microsoft Kinect) | Time Series | Physical | Sensor |
| | Student skin temperature and electrodermal activity | Time Series | Physiological | Sensor |
| Liu et al., 2019. | Speech between students | Audio | Physical | Microphone |
| | Student interaction with the system interface | Video | Physical | Webcam |
| | Student interactions with the teacher | Video | Physical | Webcam |
| | Student activities | Text | Digital | Platform |
| | Evaluation records | Numerical | Digital | Record in CSV |
| Nam Liao et al., 2019. | Student pre-requisites | Numerical | Digital | Platform |
| | Multiple-choice questionnaires | Numerical | Digital | Platform |
| | Individual student tasks | Numerical | Digital | Platform |
| Yue et al., 2019. | Facial expressions | Video | Physical | Webcam |
| | Eye movement | Time Series | Physical | Sensor |
| | Open-source image dataset for learning Asian faces | Image | Digital | Open source |
| | Dynamic mouse records for performance analysis | Time Series | Digital | Log |
| | Student scores | Numerical | Digital | Log |
| Di Mitri et al., 2017 | Leg step count and Heart rate | Time Series | Physical | Sensor |
| | Weather condition | Time Series | Environmental | Platform |
| Sharma et al. 2019 | Eye movement | Time Series | Physical | Sensor |
| | Motivation from questionnaire | Numerical | Psychometric | Platform |
| | Student scores | Numerical | Digital | Log |
| Hussain et al. 2011 | Electrocardiogram | Time Series | Physiological | Sensor |
| | Facial electromyogram | Time Series | Physiological | Sensor |
| | Respiration | Time Series | Physiological | Sensor |
| | Galvanic Skin response | Time Series | Physiological | Sensor |
| | Facial expressions | Video | Physical | Webcam |
| | Self-report affective state | Numerical | Psychometric | Platform |

In online education there was also variation between the different studies. The work presented in Wu et al. (2020) stood out for focusing on the teacher and for analyzing teacher gestures, behavior, and especially body position/pose, as well as proposing a general model of human joint positions which they used to model teacher movement in an open online classroom. Some studies, such as Brodny (2017), fused various video sources and student data from the platform. The study by Peng & Nagao (2021) stood out for the wide range of different sources, including heart rate and mental states via video and text; Luo et al. (2020) also fused video and text data, including data about posture, facial expressions, and thoughts. In Henderson et al. (2019b), the researchers fused data about posture and electrodermal activity, this time in a fully online environment (the same authors have also looked at in-person settings). The study by Liu et al. (2019) had a wide range of different types of data with numerous means for capturing it, whereas Nam Liao et al. (2019) went in the opposite direction, as it only included numerical digital data. Yue et al. (2019) used a wide range of fused data (facial expressions, eye tracking, grades, etc.) and was the only study to include an open data source in the fusion. Di Mitri et al. 2017 is the only work that used learning environment data such as temperature, pressure, precipitation, and weather type together with heart rate and step count in self-regulated learning. Finally, two works used psychometrics data. Sharma et al. (2019) proposed stimuli-based gaze analytics to enhance motivation and learning in MOOCs while the student's eye-movements were recorded. They also used a motivation scale from a 5-point Likert questionnaire. Hussain et al. (2011) detect learners' affective states from multichannel physiological signals (heart activity, respiration, facial muscle activity, and skin conductivity) during tutorial interactions with AutoTutor, an ITS with conversational dialogues. They also asked learners to provide self-reports of affect based on both categorical and dimensional (valence/arousal) models.

## 2.3 Hybrid and Blended Classroom

This section presents the different data fused from hybrid and blended learning environments. Both types of learning involve a mix of in-person and online learning, but the who differs in the two scenarios. With hybrid learning, the in-person learners and the online learners are different individuals. With blended learning, the same individuals learn both in person and online. Table 3 show the papers that used these types of data for fusing them, using the same structure as previous sections.

**TABLE 3. Sources of multimodal data in hybrid and blended settings.**

| Study | Source | Type | Category | Capture method |
|---|---|---|---|---|
| Chango et al., 2021a, 2021b. | Theory classes (attendance, attention, and notetaking) | Video | Physical | Webcam |
| | In-person practical classes (attendance and scores) | Video | Physical | Webcam |
| | Online student interactions with the platform | Numerical | Digital | Platform |
| | Final exam score | Numerical | Digital | Platform |
| Xu et al., 2019. | Classes given by the teacher | Video | Physical | Webcam |
| | Teacher speech | Text | Digital | Log |
| | Questions asked in class | Text | Digital | Log |
| Chen et al., 2014. | Head position | Video | Physical | Webcam |
| | Gaze tracking | Video | Physical | Webcam |
| | Facial expression | Video | Physical | Webcam |
| | Student electrodermal activity | Time series | Physiological | Sensor |
| | Student evaluation (attempts to answer questions, correct/incorrect responses, and final score) | Text | Digital | Log |
| Bahreini et al., 2016. | Facial features to detect emotions | Video | Physical | Webcam |
| | Vocal features to detect emotions | Audio | Physical | Microphone |
| Li et al., 2020. | Teacher/demonstrator body movement (Kinect) | Time series | Physical | Sensor |
| | Teacher/demonstrator joint positions (Myo armbands) | Time series | Physical | Sensor |
| Qu et al., 2021. | Classroom teaching data (performance, exam results) | Numerical | Digital | Log |
| | Online teaching data (performance, exam results) | Numerical | Digital | Log |
| | Offline teaching data (performance, exam results) | Numerical | Digital | Log |
| Shankar et al., 2019. | Digital tool adaptors | Numerical | Digital | CSV |
| | IoT adaptors | Numerical | Digital | CSV |

In hybrid or semi-in-person education, the work by Chango et al. (2021a, 2021b) stands out for the fusion of different types of class recordings with data obtained through Moodle, while Xu et al. (2019) fused video and text of the teacher both explaining various ideas in class and answering students' questions. The study by J. Chen et al. (2014) stood out by including probably the greatest number and widest variety of data sources to fuse, including posture, gaze, electrodermal activity, and student evaluation data. In contrast, Bahreini et al. (2016) performed emotion detection from the fusion of video of student faces and recordings of their voices. The study by Li et al. (2020) was innovative in the EDM/LA field, recording the body movement of the teacher/demonstrator using sensors and armbands, starting with a model of a human being emulated by the movement of a robot. Finally, Qu et al. (2021) fused different numerical student performance data, presenting little variety of data types in the study. Numerical data were also fused in Shankar et al. (2019), although in this case from what the authors called adaptors, one of which gathered data from the students' digital environments, while the other—the IoT adaptor—collected data from sensors physically located in the learning environment. That makes this study a good example of a hybrid using a physical-digital fusion.

## 3. DATA FUSION TECHNIQUES IN MULTIMODAL LA/EDM

This section aims to analyze the fusion process of multimodal educational data. In the next subsection we describe three fundamental aspects of this process: when the fusion is done or fusion point, what are the most used data fusion techniques, and in which EDM/LA applications/objectives data fusion has been used more.

### 3.1 When fusion is done

Data fusion techniques can be characterized in different ways depending on the area of application. Figure 3 shows the most widespread—practically standardized—categorization of MLA data fusion.

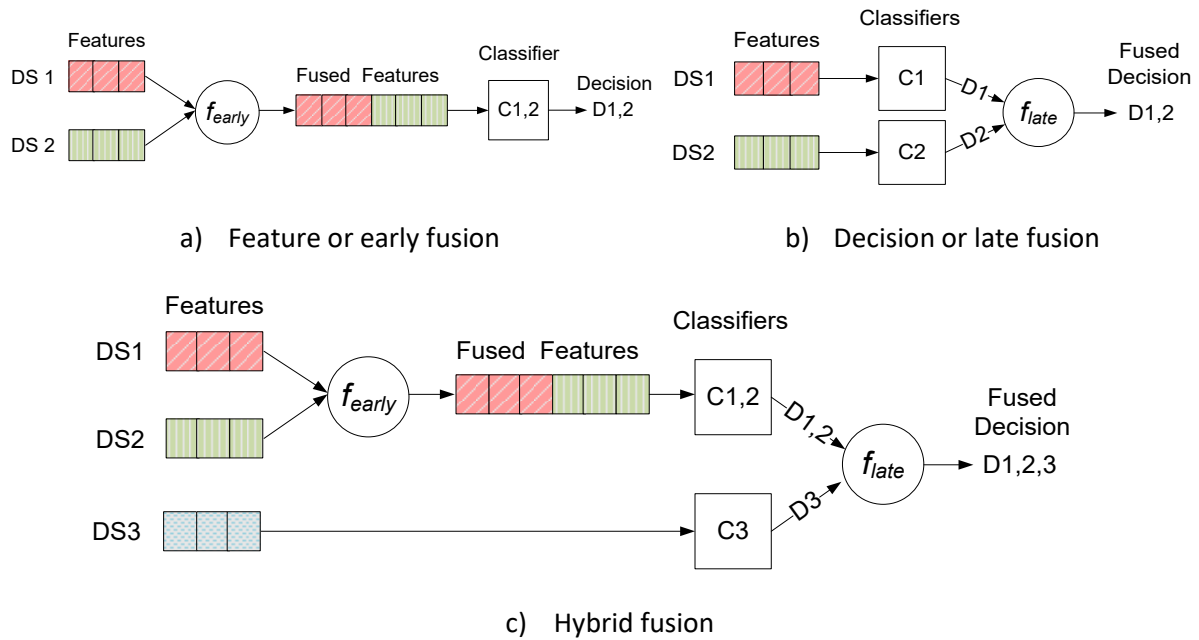

a) Feature or early fusion      b) Decision or late fusion

c) Hybrid fusion

**FIGURE 3. Multimodal data fusion schema according to when fusion is done.**

As we can see in Figure 3, data fusion techniques can be classified based on when the fusion is done, giving rise to the three following three main types (Ding et al., 2019):

- **Feature-level** or **early fusion**: a fusion approach consisting of concatenating the various features of the data from the different sources in a single vector of heterogeneous elements.

- **Decision-level** or **later fusion**: a fusion approach which consists of first creating a classifier with each of the data sources separately in order to subsequently fuse the prediction offered by the different classifiers.

- **Hybrid fusion**: a fusion approach which uses the two approaches above in a single fusion process.

Table 4, categorize the selected papers by the fusion point or the moment in which the fusion is done (early, later and hybrid fusion). It is important to note that some papers may appear in multiple categories due to they have used different time points where fusion was done. We also found some studies that don't fit into none of those three groups or in which the fusion point was not specified (Others category).

**TABLE 4. Categorization of papers by fusion point.**

| Fusion point | Explanation | Studies |
| --- | --- | --- |
| Early | Concatenation of the features of the different data sources | Andrade et al., 2016; Bahreini et al., 2016; Chango et al., 2021a, 2021b; Gadaley et al., 2020; Giannakos et al., 2019; Henderson et al., 2019a, 2019b, 2020; Liu et al., 2019; Mao et al., 2019; Nam Liao et al., 2019; Olsen et al., 2020; Peng & Nagao, 2021; Prieto et al., 2018; Shankar et al., 2019; Wu et al., 2020; Xu et al., |

| | | 2019; Yue et al., 2019; Di Mitri et al., 2017; Sharma et al. 2019; Hussain et al. 2011. |
|---|---|---|
| Later | Fusion of the predictions of each classifier (each created from a data source) | Chango et al., 2021a, 2021b; Chen et al., 2014; Daoudi et al., 2021; Peng & Nagao, 2021; Wu et al., 2020; Henderson et al., 2019a; Henderson et al., 2019b; Monkaresi et al. 2017. |
| Hybrid | A mix of the two approaches above | Brodny, 2017; Chango et al., 2021a, 2021b; Luo et al., 2020. |
| Others | Approaches that do not fit within the three described above | Li et al., 2020; Qu et al., 2021; Worsley, 2014.; Ma et al., 2015. |

Five of the studies which appeared in the early fusion category stand out for going beyond simple concatenation of features with rather more detailed procedures. Four of those studies were configured to select the best features of each data source (Chango et al., 2021a, 2021b; Henderson, et al., 2019b; Henderson et al., 2020). In contrast, Henderson et al. (2019a), reduced the dimensionality of the features using Principal Component Analysis (PCA) in two different configurations: a) they concatenated all of the features of the sources and applied PCA to the resulting vector; b) they applied PCA to the features of each source first and concatenated the results following the reduction of dimensionality. Yue et al. (2019) selected the best features first and then reduced dimensionality using two approaches, PCA and a Kolmogorov-Smirnov test. The other studies in the early category based fusion on mere concatenation of the features extracted from each source into a single vector of features which fed into the subsequent analysis.

There were also a number of studies in the later or decision fusion category, based on fusing the predictions made by the different classifiers constructed from the different data sources. Three studies used the "ensembles" approach to fuse the decisions (Chango et al., 2021a, 2021b; Wu et al., 2020). There were also four studies which based decision fusion on the decision made by the classifier with the best predictive ability (Henderson et al., 2019a; Henderson et al., 2020; Peng & Nagao, 2021; Monkaresi et al. 2017). The fusion in Henderson et al. (2019b) was done by consolidating partial decisions from each classifier into a single value, whereas the result of the fusion in Daoudi et al. (2021) was produced by weighting each classifier's decision. Chen et al. (2014) did not specify the details of their decision fusion, merely indicating that it was done. Finally, Monkaresi et al. (2017) used individual channel base classifier to make a classification by using the decision of whichever base classifier had the highest decision probability.

Various studies appeared in both the early and decision categories because the researchers made comparisons between the two approaches to determine which gave the best results. In contrast, only three studies appeared in the hybrid fusion category, using early and decision fusion in combination. Brodny (2017) proposed a conceptual model with feature fusion at the beginning followed by decision fusion, although it was laid out very broadly. In addition to the basic schemes of early and decision fusion described above, Chango et al. (2021a; 2021b) also used hybrid configurations at some point in which the features of some sources were fused at the beginning to produce classifiers which were subsequently fused by means of ensembles. This study also stood out for the richness of the experiments, as it compared the hybrid approach with purely early and late fusion approaches.

Finally, four studies did not fit in any of the previous categories. Li et al. (2020) used a fusion model of sensors from various data sources in order to produce more accurate data during a demonstration by an instructor for a robot to learn to do certain tasks. Qu et al. (2021) proposed a five-step fusion process which affected the features produced by the data sources via a weighting technique but which did not fit within the early fusion category because there was no concatenation of features as such. The work presented in Worsley (2014) was a general article proposing three generic fusion approaches which differed from the traditional early-decision-hybrid categorization. In contrast, the author talked about naïve fusion (equivalent to feature fusion), low-level fusion (where the researcher is intentional about enacting multimodal data fusion on very small-time scales because they may have prior knowledge that the various modalities have time-specific relevance to one another), and high-level fusion (which takes the data to a higher level of meaning, equating features to states). Ma et al. (2015) produced a process model of multisource data fusion analysis and, according to the authors, put it into practice from the dimension of data fusion. There is only one paragraph in the article explaining the fusion, but it is so abstract that it is impossible to determine when the fusion was done or what data was affected.

## 3.2 Fusion technique

We have classified the selected papers by the used fusion technique in the main types or categories (see Table 5). The criteria for classification were the fundamental data fusion techniques used, from the purist perspective of data fusion: aggregation, ensembles, statistical operators, mathematical operators, similarity-based, probability and filters. Again, the categories were not exclusive as some studies used more than one technique. There were also studies which did not fit into the standard data fusion categories and studies which did not specifically state the fusion technique used (Others in the table).

**TABLE 5. Categorization of papers based on fusion technique.**

| Technique | Explanation | Studies |
|---|---|---|
| Aggregation | Fusion consists of aggregating (in the sense of concatenating) data from the different sources | Andrade et al., 2016; Bahreini et al., 2016; Chango et al., 2021a, 2021b; Gadaley et al., 2020; Giannakos et al., 2019; Henderson, 2019a, 2019b, 2020; Liu et al., 2019; Nam Liao et al., 2019; Olsen et al., 2020; Peng & Nagao, 2021; Prieto et al., 2018; Shankar et al., 2019; Worsley, 2014; Wu et al., 2020; Xu et al., 2019; Yue et al., 2019; Di Mitri et al., 2017; Sharma et al. 2019; Hussain et al. 2011. |
| Ensembles | Applies the idea of ensembles (machine learning) to combine the data from the various classifiers' decisions | Chango et al., 2021a, 2021b; Wu et al., 2020. |

| Statistical operators | Uses statistical operators to combine the data from the different sources | Daoudi et al., 2021; Henderson et al., 2019a, 2020; Qu et al., 2021. |
|---|---|---|
| Mathematical operators | Uses mathematical operators to combine the data from the different sources | Qu et al., 2021. |
| Similarity-based | Fusion is based on the calculation of similarities | Qu et al., 2021. |
| Probability | The fusion uses the concept of probability, normally linked to the concept of "certainty" provided by each data source. | Peng & Nagao, 2021; Monkaresi et al. 2017. |
| Filters | Data are fused via the use of filters, generally used for estimating the hidden state of a dynamic system | (Li et al., 2020) |
| Others | Non-standard fusion techniques which do not fit in any of the other categories | (Luo et al., 2020)(N. L. Henderson, Rowe, Mott, Brawner, et al., 2019) (Brodny, 2017)(J. Chen et al., 2014)(Mao et al., 2019)(Ma et al., 2015) |

Perhaps the most elementary data fusion technique consists of combining data in the most basic sense of aggregating or concatenating. A large number of studies fell within this category because it is specifically the idea of concatenation of data which underlies early fusion studies in which features are aggregated or concatenated. Work which stood out in this category included Worsley, (2014) for the terminology used, which associated the term "naïve" fusion with aggregation (of features in this case); Prieto et al. (2018), for the aggregation of data from wearables, with the peculiarities that went with it, and the various studies by Henderson et al. (2019a, 2020) who demonstrated a consistent line of research in this type of educational data fusion.

Ensembles were also an interesting fusion approach, combining the results offered by various classifiers. This category included studies such as Wu et al. (2020), who used the "stacking" method, and Chango et al. (2021a, 2021b) who used the "vote" approach provided by the Weka tool. Details of both approaches may be found in each of the articles cited.

Another common way of combining data was to calculate some statistic from the data which summarized it. Daoudi et al. (2021) calculated the weighted means of the different classifiers constructed in the decision fusion process. Henderson et al. (2019a, 2020) calculated the maximum value from the values for predictive reliability from the various classifiers. Qu et al. (2021) used the Spearman coefficient as a measure of correlation between the data sources during the fusion process.

That same study by Qu et al. (2021) appears in the category of fusion methods based on mathematical operators and the category of those based on similarity because, after calculating the Spearman coefficient, the authors used various mathematical formulas to weight the data sources, including some which used the concept of similarity. That means that this study has an advanced, rather than elementary, fusion approach which might open the door for other researchers to not restrict themselves to the classical aggregation of features.

Probabilistic theory is also applicable in data fusion. We found only one of the studies in this category, Peng & Nagao (2021), who chose the classifier in a late fusion process according to the probability distributions of each of the classifiers being correct. It is interesting to note that in this study the authors also spoke of a type of fusion called "single-channel level" fusion, which in our opinion does not actually refer to any kind of fusion as it is based on just the construction of classifiers from each of the data sources. Monkaresi et al. (2017) used individual channel base classifier to make a classification by using the decision of whichever base classifier had the highest decision probability.

In data fusion, filters are one of the most well-known and widely applied approaches as they allow environmental data to be fused in order to predict future states in dynamic systems. Despite that, we only found one study falling within this category (Li et al., 2020), which used the Kalman filter to fuse data of different modalities and the dynamic time warping algorithm to align the data in the same timeline. The idea was that the fused data, coming from sensors placed on an instructor's body during a demonstration of a task, would allow a robot to predict the actions it needed to do to imitate that as faithfully as possible.

Two of the studies used other non-standard fusion techniques. There was a very elaborate fusion strategy in Luo et al. (2020). On the one hand, the features obtained from student interaction data on the platform were weighted according to their entropy and then fused by means of aggregation using that weighting, constructing a first classifier from those features which reflected the students "thinking" aspect. Subsequently, using video recordings, two classifiers were created reflecting "attention" and "emotion" aspects, with those two classifiers being fused at the decision feature level using the Analytic Hierarchy Process (AHP) technique. That fused decision of attention and emotion was then fused with the thinking classifier at the decision level using AHP. This is summarized in Figure 4. Henderson et al. (2019b) used a feature fusion configuration via aggregation of features (both in the most basic version and in the more advanced version which selected the best features) and, the most interesting aspect of the article, the decision fusion configuration used the Match-score fusion technique (Rahman & Gavrilova, 2018).

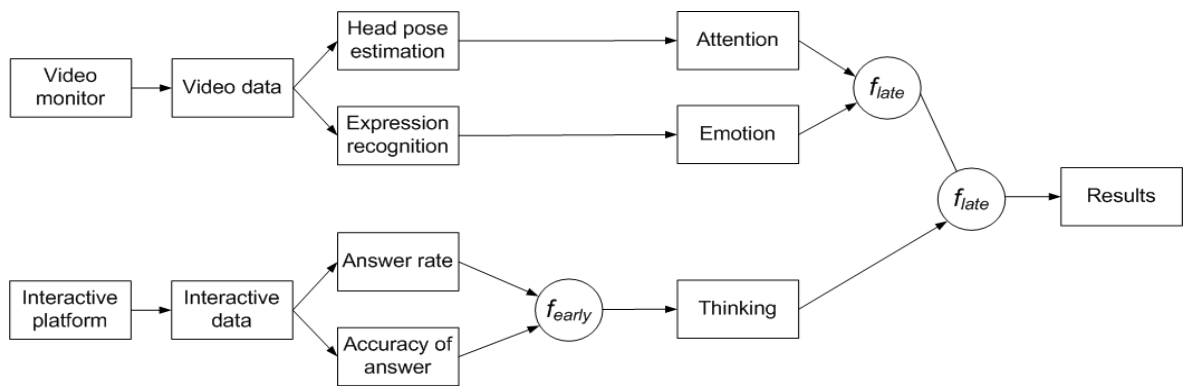

**FIGURE 4. Example of the advanced fusion approach used in Luo et al. (2020).**

It was not possible to determine the fusion techniques used in some studies. Despite Brodny (2017) proposing a hybrid fusion technique, it was in broad conceptual terms without any specific fusion techniques being mentioned, it lacked specific test results or tangible results. Chen et al. (2014) indicated that their study dealt with decision fusion but did not offer specific details about the fusion approach used. The title of Mao et al. (2019) indicated that there was multi-feature fusion but the article did not give details of the features or the fusion method used. Finally, the study by Ma et al. (2015), from which, as already noted, it was not possible to determine the exact point of fusion, also failed to provide information to identify the data fusion technique used.

## 3.3 EDM/LA objective

We have classified the selected papers based on the EDM/LA objective or education application that want to resolve the paper that use multimodal data fusion. There are a wide range of popular application or objectives in LA/EDM (Romero and Ventura, 2013) (Romero and Ventura, 2020) for solving educational problems or goals. Table 6 shows the categorization of the papers based on commonly sought objectives in the areas of EDM and LA. There were also other different objectives and some papers which did not specifically indicate the EDM/LA objective (Others in the table).

**TABLE 6. Categorization of papers according to EDM/LA objective.**

| EDM/LA Objective | Explanation | Studies |
|---|---|---|
| Analysis of students' learning processes | To analyze the student's behavior and style during learning and discovering patterns. | Andrade et al., 2016; Liu et al., 2019; Ma et al., 2015; Qu et al., 2021; Shankar et al., 2019. |
| Prediction of students performance | To infer the students final performance/mark/grade variable from some combination of other variables. | Chango et al., 2021a, 2021b; Giannakos et al., 2019; Nam Liao et al., 2019; Olsen et al., 2020; Di Mitri et al., 2017. |
| Students' emotional state evaluation/recognition | To study affect during learning and the importance of students' emotions to learning. | Bahreini et al., 2016; Brodny, 2017; Chen et al., 2014; Daoudi et al., 2021; Henderson et al., 2019a, 2019b, 2020; Mao et al., 2019; Peng & Nagao, 2021; Hussain et al. 2011. |
| Prediction of students' engagement | To predict students' engagement, motivation, interest, etc. | Gadaley et al., 2020; Luo et al., 2020; Yue et al., 2019.; Monkaresi et al. 2017; Sharma et al. 2019. |
| Modelling teacher behaviour | To analyze teacher behavior during instruction and interaction with students. | Prieto et al., 2018; Wu et al., 2020. |
| Teacher discourse classification | To analyze instructors' text data from forums, chats, social networks, etc. | Xu et al., 2019. |
| Others | Other objectives or applications | Li et al., 2020.; Worsley, 2014. |

There was a group of studies in which the objective was to conduct analyses of students' learning processes. Liu et al. (2019) sought to understand student learning processes, incorporating insights from data collected in multiple modalities and contexts. Ma et al., (2015) aimed to analyze the learning process in a smart classroom. Qu et al. (2021) evaluated college students' learning behavior, providing a basis for adaptive learning environments. The objective for Shankar et al. (2019) was to better understand the learning process considering the contextual information of the situation. Andrade et al. (2016) modelled student behavior in order to identify whether clusters of observable behaviors could be used to identify and characterize behavioral frames in rich video data of student interviews.

Other studies sought to predict students' final performance, such as López-Zambrano et al. (2021). Chango et al. (2021a, 2021b) predicted the final academic performance of university students in a blended learning environment. In contrast, Giannakos et al. (2019) sought to accurately predict users' acquisition of skills, commonly called movement-motor learning. The aim in Olsen et al. (2020) was to predict students' collaborative learning gains, while in Nam Liao et al. (2019), it was to arrive at early predictions of students' overall performance on a course. Di Mitri et al. (2017) uses a machine learning approach for predicting performance in self-regulated learning. Sharma et al. (2019)

There was a group of studies in which fusion was used to try and improve the evaluation of emotions during the learning process. The main objectives sought in this regard were: the evaluation of learners' affective states in collaborative serious games (Daoudi et al., 2021); emotion recognition and integration of emotional states in educational applications with consideration of uncertainty (Brodny, 2017); recognizing student mental states in conversations (Peng & Nagao, 2021); emotion recognition (frustration) in game-based learning (Henderson et al., 2019b); recognition of students' affective states (Chen et al., 2014); recognition of students' facial micro expressions (Mao et al., 2019); real-time, continuous, unobtrusive emotion recognition (Bahreini et al., 2016); improved detection of affect (Henderson et al., 2020); detection of learner affect in game-based learning (Henderson, 2019a); and (Hussain et al. 2011) detect learners' affective states in ITS (Intelligent Tutoring Systems).

In other studies, fusion was used to improve predictions of student engagement or interest. Gadaley et al. (2020) predicted engagement in classes where students were allowed to have digital devices during lectures. Luo et al. (2020) modelled student interest using multimodal natural sensing technology in order to provide an effective basis for improving teaching in real time. The aim in Yue et al. (2019) was to detect learners' emotional and eye-based behavioral engagement in real-time as well as to predict learners' performance after completing a short video course. Monkaresi et al. (2017) explored how computer vision techniques can be used to detect engagement while students completed a structured writing activity (draft-feedback-review). Sharma et al. (2019) proposed stimuli-based gaze variables as student's attention indicators (i.e. with-me-ness) in order to enhance motivation and learning in MOOCs.

One group of studies was notable because they modeled teacher behavior, with fusion being used to assist that modelling. The aim in Wu et al. (2020) was to recognize teacher behavior in order to solve problems of time-consumption and information overload in teaching and then help teachers optimize teaching strategies and improve teaching efficiency. The objective in Prieto et al. (2018) was pedagogical modelling of a teacher in class in order to provide automated tagging of classroom practice that could be used in everyday practice with multiple teachers. Finally, there was one study in which fusion helped to classify teacher speech, (Xu et al., 2019). The aim was to automatically classify teacher discourse in a Chinese classroom. One intriguing objective (in the "others" category) was the automatic reproduction of learned tasks by a robot following a teachers' demonstrations of certain physical tasks (Li et al., 2020). Lastly, there was one study in which the author indicated that they did LA, but did not specify any LA objectives, simply discussed how decisions about data fusion have a significant impact on how the research relates to learning theories (Worsley, 2014).

We have also conducted an analysis to reveal the relationship between the data fusion technique used in the different works and the EDM/LA objectives achieved in each of those work (see Table 7).

**TABLE 7. Relationship between Fusion technique and EDM/LA objective.**

| Technique / Objetive | Aggregation | Ensembles | Statistical operators | Math operators | Similarity-based | Probability | Filters | Others |
|---|---|---|---|---|---|---|---|---|
| **Analysis of students' learning processes** | Andrade et al., 2016; Liu et al., 2019; Shankar et al., 2019. | | Qu et al., 2021. | Qu et al., 2021. | Qu et al., 2021. | | | Ma et al., 2015. |
| **Prediction of students performance** | Chango et al., 2021a, 2021b; Giannakos et al., 2019; Nam Liao et al., 2019; Olsen et al., 2020; Di Mitri et al., 2017. | Chango et al., 2021a, 2021b. | | | | | | |
| **Students' emotional state evaluation/recognition** | Bahreini et al., 2016; Henderson, 2019a, 2019b, 2020; Peng & Nagao, 2021; Hussain et al., 2011. | | Daoudi et al., 2021; Henderson et al., 2019a, 2020. | | | Peng & Nagao, 2021. | | N. L. Henderson, Rowe, Mott, Brawner, et al., 2019; Brodny, 2017; J. Chen et al., 2014; Mao et al., 2019. |
| **Prediction of students' engagement** | Gadaley et al., 2020; Yue et al., 2019; Sharma et al., 2019. | | | | | Monkaresi et al., 2017. | | Luo et al., 2020. |
| **Modelling teacher behaviour** | Prieto et al., 2018; Wu et al., 2020. | Wu et al., 2020. | | | | | | |
| **Teacher discourse classification** | Xu et al., 2019. | | | | | | | |
| **Others** | Worsley, 2014. | | | | | | Li et al., 2020 | |

If we analyze the above table from the perspective of data fusion techniques, we can see that aggregation techniques cover all the range of EDM/LA objectives, having most works located on the categories of analysis of students' learning process, prediction of performance, emotional recognition and engagement prediction, which shows the wide applicability of aggregation-based fusion approaches. From the perspective of EDM/LA objective, we have noticed that analysis of students' performance and emotional state recognition are two objectives covered by a large number of different fusion approaches (five and four respectively), having found that aggregation approach is the most used when trying to achieve those two objectives. When it comes to prediction of students' performance, aggregation and ensembles are the two only approaches employed, which indicates that both seems to be the reference standard approaches for achieving those two objectives. Focusing on engagement prediction and students' behavior modeling, aggregation-based fusion also is the most employed approach. Finally, emotional state recognition is an objective for which the use of aggregation, statistical operators and other non-standard approaches have demonstrated to obtain positive results.

## CONCLUSIONS AND FUTURE TRENDS

Data Fusion of multimodal data seems promising in the field of Education in general (Sultana et al., 2020), and particularly in the field of EDM/LA (Mu et al., 2020), as we have shown in this review. We have analyzed all the related papers in the bibliography from the perspective of the data being fused, the fusion approaches used, and the EDM/LA educational objective or application and we have obtained the next conclusions:

- In terms of the data fused, there was a relatively balanced use in the different educational environments, with data fusion being found 12 papers focused on in-person learning, 11 on online learning, and 8 on hybrid and blended environments. Most of the data being fused are focused specifically on learners, while only a minority focused on teacher data. The data came from a wide range of sources, mainly recordings of students, sensor readings of various aspects, and numerical data indicating some magnitude generally related to academic performance. Almost all of the data were physical or digital, a minority were physiological.

- In terms of the fusion approach, the majority of the papers used early fusion of features, while a large number used late fusion or decisions produced by different classifiers in previous stages. Very few studies used hybrids of those two approaches and even fewer went outside this framework (early-late-hybrid) summarized in Figure 2. Looking at the fusion techniques used, aggregation of features is the predominant method, followed by others based on the use of statistical operators and ensembles.

- In terms of EDM/LA objectives or educational application/problem in papers that used data fusion, the most notable were those seeking to manage student emotions, analyze student behavior, and those that aimed to predict academic performance, interest, or engagement.

It is important to notice that only one paper (Olsen et al., 2020) used a free available public multimodal dataset from Pittsburgh Science of Learning Center (PSLC) DataShop. All the other selected papers used their own private datasets. However, there are an increasing number of Publicly Available EDM/LA Datasets (Mihaescu and Popescu, 2021) and Table 8 shows a list of specific multimodal datasets/repositories that could be used for researching on EDM/LA data fusion.

**TABLE 8. Publicly available EDM/LA multimodal datasets.**

| Name | URL |
| --- | --- |
| Dataset of Multimodal Interface for Solving Equations | https://pslcdatashop.web.cmu.edu/Project?id=33 |
| MUTLA: A Large-Scale Dataset for Multimodal Teaching and Learning Analytics | https://paperswithcode.com/dataset/mutla |
| NUS Multi-Sensor Presentation (NUSMSP) Dataset | https://scholarbank.nus.edu.sg/handle/10635/137261 |
| PE-HRI: A Multimodal Dataset for the study of Productive Engagement in a robot mediated Collaborative Educational Setting | https://zenodo.org/record/4288833#.Yd4OO_7MKUk |
| Student Life Dataset | https://studentlife.cs.dartmouth.edu/ |
| VLEngagement: A Dataset of Scientific Video Lectures for Evaluating Population-based Engagement | https://github.com/sahanbull/context-agnostic-engagement |

Finally, after doing this review of the literature about data fusion of multimodal data, we identify the next opportunities or challenges for future research in this area:

- Most of the data sources examined were from students, with only a few studies focusing solely on teachers. It would be interesting to combine both teacher and student data in the same study in order to determine whether student behavior could be influenced by teacher characteristics, or whether the teacher adapts their methodology based on the type of student they are teaching, such as in the framework of the classical theories from Biggs (1987) about student and teacher approaches to learning.

- We found only some works about fusion of psychometric/environmental data in the processes we examined. It would be interesting to see more works using this type of data to, for example, be able to determine whether student psychological processes are affected in any way by the nature of the environment in which they are learning (temperature, humidity, lighting, etc.).

- Most fusion techniques used in EDM/LA are basic and fundamental, and the most widely used are simple aggregation, ensembles, and statistical operators. It is clear that is restricted to early-late-hybrid fusion schema. However, data fusion is such a rich, versatile field that much potential is lost by that restriction. Some studies proposed using other, more flexible processes, which produced good results in some studies and should be explored (Li et al., 2020; Qu et al., 2021; Worsley, 2014). Other advanced approaches that allow improved fusion in different fields are: the use of techniques based on filters, probabilistic approaches, possibilistic approaches, and the use of the Demptster-Shafer theory of evidence seem useful for that and have been little used for educational data fusion.

- Data fusion of multimodal data has been used mainly in some EDM/LA applications such as prediction of performance and engagement, analysis of student behavior, and the management of student emotions. However, there are much other EDM/LA objectives, applications or educational problems that have not been addressed by using data fusion such as classroom planning, learning strategy recommendations, and course construction and organization, etc.

- It is also important to mention the connection between multimodal data fusion to the knowledge management (KM) area and incorporate its large experience and vast literature data management process. In this line, it would be interesting to see works showing for example: how data fusion would be benefitted in cloud-based knowledge management frameworks in higher education institutions (Noor et al., 2019) or in visualizing educational information in the context of modern education megatrend (Izotova et al. 2021).

## ACKNOWLEDGMENTS

This work would not have been possible without the funding from the Ministry of Sciences and Innovation I+D+I PID2020-115832GB-I00 and PID2019-107201GB-I00 and funding for open access charge: Universidad de Córdoba / CBUA.